\documentclass[12pt]{article}

\usepackage[utf8]{inputenc}
\usepackage{charter}
\usepackage{fullpage}
\usepackage{hyperref}
\usepackage{graphicx}
\usepackage{lipsum}
\usepackage{tabularx}
\usepackage[sort&compress,numbers]{natbib}
\usepackage{hyperref}
\hypersetup{
    colorlinks=true,
    linkcolor=blue,
    filecolor=blue,
    citecolor=blue,
    urlcolor=blue,
}
\usepackage[total={6.5in,9in},top=1in,headsep=0.1in,headheight=1in]{geometry}

\newcommand\snowmass{
\begin{center}
  \rule[-0.2in]{\hsize}{0.01in}\\
  \rule{\hsize}{0.01in}\\
  \vskip 0.1in
  Submitted to the Proceedings of the US Community Study\\ 
  on the Future of Particle Physics (Snowmass 2021)\\
  \rule{\hsize}{0.01in}\\
  \rule[+0.2in]{\hsize}{0.01in}\\[-2em]
\end{center}
}

\usepackage[firstpage=true]{background}
\backgroundsetup{contents={\parbox{6.5in}{\snowmass}}, scale=1,placement=top,opacity=1,color=black,position={3.25in,1.2in}}

\usepackage{fancyhdr}
\fancypagestyle{plain}{%
  \fancyhf{}%
  \fancyhead[C]{}
  \fancyfoot[C]{\thepage}
}

\fancypagestyle{empty}{%
  \fancyhf{}%
  \fancyhead[C]{{\it Snowmass2021 Accelerator Frontier White Paper: Near Term Applications driven by Advanced Accelerator Concepts}}
  \fancyfoot[C]{\thepage}
}
\title{Snowmass2021 Accelerator Frontier White Paper: Near Term Applications driven by \\ Advanced Accelerator Concepts}
\date{\today}

\usepackage{authblk}

\author[1]{Claudio Emma}
\author[2]{Jeroen van Tilborg}
\author[3]{Félicie Albert}
\author[4]{Luca Labate}
\author[1]{Joel England}
\author[1]{Spencer Gessner}
\author[1]{Frederico Fiuza}
\author[2]{Lieselotte Obst-Huebl}
\author[5]{Alexander Zholents}
\author[6]{Alex Murokh}
\author[7]{James Rosenzweig}

\affil[1]{SLAC National Accelerator Laboratory, Menlo Park, California 94025, USA}
\affil[2]{Lawrence Berkeley National Laboratory, Berkeley, California 94720, USA}
\affil[3]{Lawrence Livermore National Laboratory, Livermore, California 94550, USA}
\affil[4]{Istituto Nazionale di Ottica (INO), Consiglio Nazionale delle Ricerche (CNR), 56124 Pisa, Italy}
\affil[5]{Argonne National Laboratory, Lemont, Illinois 60439, USA}
\affil[6]{RadiaBeam Technologies, Santa Monica, California 90404, USA}
\affil[7]{University of California — Los Angeles, Los Angeles, California 90095, USA}

\begin{document}

\maketitle

\begin{abstract}

While the long-term vision of the advanced accelerator community is aimed at addressing the challenges of future collider technology, it is critical that the community takes advantage of the opportunity to make large societal impact through its near-term applications. In turn, enabling robust applications strengthens the quality, control, and reliability of the underlying accelerator infrastructure.
The white paper contributions that are solicited here will summarize the near-term applications ideas presented by the advanced accelerator community, assessing their potential impact, discussing scientific and technical readiness of concepts, and providing a timeline for implementation.

\end{abstract}
\newpage
\tableofcontents
\newpage
\section{Executive Summary}

In recent years, an abundance of applications of compact particle accelerators and their associated radiation sources (electron, protons, X-rays, gammas, etc.) has emerged. Not only do advanced concepts shrink down the accelerator size, and cost, but accessing new frequencies and time-scales in accelerating regimes can enable new types and properties of radiation pulses.
Not surprisingly, in an ever-more technological world, controlling and diagnosing matter on ultra-small and ultra-short scales is the key driver of innovation and discovery. 
For example, the 2020 Basis Research Needs (BRN) report for ``Transformative Manufacturing" \cite{BRN-TransformativeManufacturing-2020} lists accelerators and X-ray light source as critical in novel material processing and characterization. The BRN report ``Microelectronics" \cite{BRN-MicroElectronics-2018} emphasized advanced accelerator needs, by citing ``{\it ..how functional materials grow, nanoscale imaging, provided by these x-ray techniques, is an invaluable tool to characterize high-density devices under in-operando conditions}".

Following the previous 2013 Snowmass meeting,  the relevance of near-term applications of accelerators was only discussed in the context of communication and awareness among the general public (Section 10.4.2 in Ref. \cite{Snowmass2013} states ``{\it Particle physics research has had a significant impact on other areas of science. Examples include applications developed for health and medicine...}"). This sentiment was echoed in the 2014 P5 report \cite{P5-2014} through ``{\it In return, developments within the particle physics community have enabled basic scientific research and applications in numerous other areas. This broad, connected scientific enterprise provides tremendous benefits to society as a whole.}". However, we feel that the urgency to direct the funding and efforts to near-term application development has been lacking in these previous reports. A stronger emphasis on near-term applications of advanced accelerator concepts is critical both to society in general but also to the robustness of accelerator technology itself. Dedicated funding should be channeled towards strengthening accelerator applications that fall outside of the long-range particle collider mission. A successful near-term application environment will naturally guide particle accelerator technology to maturity. Applications demand robust day-to-day performance with minimal start-up time, control over parameters such as particle flux and energy, active stabilization concepts to keep the "up-time" large throughout the day, pump-probe synchronization, and reduction in operational expenses such as energy usage and component replacements, among others.
While this was recognized in the 2016 Advanced Accelerator Development strategic report \cite{AACreport-2016}, with near-term applications on the high-energy physics roadmap, we emphasize that such language be included in the 2021 Snowmass and 2022 P5 reports. Furthermore, we emphasize that the near term applications discussed here enable opportunities for cross-cutting development in other Accelerator Frontier priorities (e.g. novel RF technology discussed in these proceedings \cite{whitepaperAF7}) as well as in the Community Engagement Frontier (e.g. industrial applications based on FLASH radiation therapy in these proceedings \cite{whitepaperFLASH}) .

While applying advanced accelerator concepts for high-quality light sources was a topic of debate just a decade ago, this white paper showcases recent impressive experimental demonstrations. For example, two plasma-accelerator milestone demonstrations of free-electron lasing \cite{Wang2021, Pompili2021b}, known to critically depend on excellent electron beam phase space characteristics, highlight recent advances.
Note that at the international level, strong competition has emerged for accelerator development for near-term applications. For example, the 2022 European particle physics report has specifically called out near-term applications (see ``Integration and Outreach" in Fig. 4.1 of \cite{EuroRoadmap2022}) as critical in achieving the long-term roadmap goals, which has led to concrete investments. The  Eupraxia project \cite{Assman2020} is a concrete example of a community-driven merging of accelerator concept and application needs. It cannot be ignored that the two aforementioned plasma-based FEL demonstrations occurred in Asian and European facilities.

In terms of the need for application-oriented facilities, the authors of this white paper recognize that the applications can be divided in two categories. For those where significant R\&D is still needed (for example, plasma-based and other advanced X-ray FEL concepts), it would be critical to obtain financial support for novel capabilities and beamlines at existing and new facilities. For those applications where proof-of-principle concepts have matured already (for example, the betatron X-ray source and laser-driven particle beams), a dedicated beamline or facility is lacking where these sources can be permanently exploited, allowing both for efficient user-access as well as dedicated photon or particle source optimization. Many of the successes to date have been realized in short, few weeks-long campaigns, followed by beamline reconfiguration for other experiments. This has hampered source optimization, stability improvements, and integration of  advanced transport and sample technology.

Table~\ref{tab:my_label} summarizes a selection of key near-term applications that are, or will be, enabled by advanced  accelerator concepts in the next 5-10 years. These span broadly across light source, medical and fundamental science applications and are described in further detail in the body of the manuscript. 


\begin{table}
    \centering
    \scalebox{0.75}{
 \begin{tabular}{|p{0.3\linewidth}|>{\centering\arraybackslash}p{0.3\linewidth}|>{\centering\arraybackslash}p{0.3\linewidth}|>{\centering\arraybackslash}p{0.3\linewidth}|}
    \hline
         \textbf{Source} & \textbf{Example application} & \textbf{Status} & \textbf{Readiness in 5-10 years}\\ \hline
        Plasma-based FEL \cite{Emma2021} & Single-shot high-res imaging, non-linear excitation & Experimental feasibility demonstrated, two high-impact papers in 2021 & Realistic at higher flux and photon energy in $<$ 5 years  \\ \hline
        Corrugated-structure FEL \cite{Zholents2020} & Medical imaging & Conceptual Development  &  Technology to be explored\\ \hline
        Cryo-cooled Copper FEL  \cite{Rosenzweig2020} & Ultrafast Imaging, Attosecond Science & Conceptual Design & Technology to be explored\\ \hline
        Betatron X-rays  \cite{Corde_RMP_2013,Albert_PPCF_2016} &  Single-shot phase-contrast imaging of micro-structures    & Extensive demonstrations     &  Ready now  \\ \hline
        Compton-scattered X-rays \cite{Corde_RMP_2013,Albert_PPCF_2016} & Compact dose-reduced medical imaging, HED dynamics &   Proof-of-principle demonstrations  & Tunable and mono-energetic in $<$ 5 years  \\ \hline
        Advanced gamma ray sources \cite{Glinec_PRL_2005,Benismail_NIMA_2011,Gadjev2019,Sudar2020} & Security, efficient imaging at reduced dose & Experimental demonstrations (plasma based).
        Conceptual Development (non-plasma based) & Plasma-based ready now. Non-plasma based technology to be explored\\ \hline
        VHEE \cite{Svendsen2021,Labate2020,Kokurewicz2019} & Low dose radiotherapy & Well established, needs stability emphasis &  Ready now at compact low rep rate sources \\ \hline
        Laser-solid ions  \cite{Linz2016}&  Medical imaging, FLASH therapy, HED diagnostic & Extensive demonstrations in TNSA regime & Ready now, $>$100 MeV protons in $<$5 years\\ \hline
        High-energy particle beams \cite{GessnerLOI} & Beam-dump explorations, astro-physical plasmas  & Initial experiments planned  & Results from initial experiments expected in $\sim$5 years\\ \hline
    \end{tabular}}
    \caption{High-level summary of near term applications of advanced accelerators described in the manuscript. References, example applications, status and readiness in 5-10 years are included for each source.}
    \label{tab:my_label}
\end{table}


\section{Introduction}

Applying advanced accelerator concepts to collider technology is driving the community to ever improve on particle beam brightness, maximum energy, and repetition rate, all in order to meet the demanding luminosity requirements. However, en route to collider applications, the particles beams and the radiation sources they drive have attractive properties for immediate and near-term applications. In fact, one could safely argue that developing near-term applications such as medical treatment and single-shot tomography, which forces challenges such as stability, reliability, tunability, and user-friendly operation to be resolved, in turn accelerates the long-term road map to particle colliders. In addition to the immediate societal benefits of near-term applications, the success of the applications requires precision and control of accelerator technology, which is an investment well worth making.

In the next three chapters we have separated the near-term applications, based on community input, into three categories: Photon/light sources, particle sources for medical applications, and fundamental high-energy physics applications.



\section{Light source applications}

\subsection{Free Electron Lasers}

High-brightness electron beams enable intense X-ray photon production in free-electron lasers (FELs) \cite{Pellegrini2016}. For example, the LCLS FEL at SLAC can deliver photons up to 25 keV energies at 120 Hz repetition rate, at a flux of $10^{11}-10^{13}$ photons per shot in 10-250 femtosecond duration pulses \cite{LCLSParameters}. Such a facility allows for single-shot imaging of complex nano-structures and biological macro-molecules in the so-called "image-before-destroy" configuration, complemented with femtosecond-resolution pump-probe capabilities for dynamics exploration. 


Due to the size (km-scale) and cost (B$\$$-scale) of the few FEL facilities over the world, the total number of experiments and applications that can be pursued is quite limited, and is prioritized by beam time committees and management priorities. FEL applications in industry (for example, transformative manufacturing \cite{BRNmanufacturing2020} and high-volume defect tomography in micro-electronics \cite{BRNmicroelectronics2018}), at smaller universities (for example for time-resolved nano-scale imaging \cite{Barty2008}), and as complementary capability to existing accelerator and light source facilities, could be opened up to users if the size and cost of the FEL facility could be significantly reduced. For this reason, a growing and active global community is exploring advanced accelerator concepts for compact FELs. 

\begin{figure}[htbp]
\centering
\fbox{\includegraphics[width=0.98\linewidth]{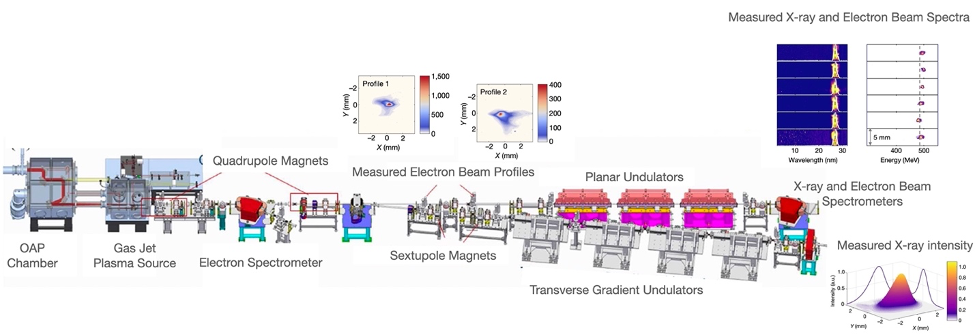}}
\caption{Figure adapted from Ref. \cite{Emma2021}. Results from the first plasma-based FEL demonstration experiment (see Ref. \cite{Wang2021} for details).}
\label{fig:wang}
\end{figure}

One promising path to compact and economic FELs involves the reliance on plasma-based accelerators. Dense relativistic electron beams or ultra-intense laser pulses can drive charge separation in meter-scale plasmas that can support accelerating fields over 2-3 orders of magnitude larger than with RF-based technology. For laser-driven wakefield accelerators (LWFAs), energies up to 8 GeV energy gain have been reported \cite{Gonsalves19} in a 20-cm plasma structure, as well as per-mille level energy spread \cite{Ke2021}, 10-100 pC charge, few-fs beam duration \cite{Buck11}, sub-mrad divergence, few-$\mu$m source size \cite{Barber17}, and repetition rates up to 10 Hz and planned for multi-kHz with laser improvement projects underway. 
For beam-driven wakefield accelerators (PWFAs), energies up to 84 GeV (42 GeV energy gain) have been reported \cite{Blumenfeld2007}, as well as per-mille-level energy spread \cite{Lindstrom2021,Pompili2021}, 10-100 pC charge, tens of fs beam duration \cite{Litos2014,Litos_2016}, $\mu$m-level normalized emittance \cite{Shpakov2021}, 10 $\mu$m  level source size, and few Hz repetition rates with plans for kHz repetition rates.
There are a growing number of experimental plasma-accelerator programs interested in using these electron beams to drive next-generation light sources. 

The community has recently produced a review paper Ref. \cite{Emma2021} summarizing the global R$\&$D efforts geared towards a plasma-based FEL. Recent highlights include the first demonstration of FEL gain at 27nm by a LWFA-driven FEL \cite{Wang2021}, and FEL gain at 820 nm by the witness beam in an PWFA FEL \cite{Pompili2021b}. Although the FEL gain is still limited in these experiments, and the FEL emission has not reached the X-ray regime, these milestone demonstrations provide evidence that the electron beam quality in plasma-based accelerators can support the phase-space quality needed for FEL lasing. Near-term efforts in the community are focusing on increasing the repetition rate, improving the shot-to-shot stability and optimizing the phase-space for FEL lasing (including application of chicanes and transverse gradient undulator to mitigate energy spread, strong-focusing undulators to improve beam density throughout the lasing process, and high-gradient transport elements to reduce facility length and to mitigate emittance growth). While fiber-based and bulk-amplifier technology has been identified, it should be stated that upgrading FEL-applicable laser technology to multi-kHz repetition rates will require steady investments over the next 5-10 years. The plasma-based compact FEL efforts, providing funding will remain to be secured, is well underway to provide  availability to users within the near-term 5-10 year framework. 


A different "advanced accelerator" approach to compact FELs has been proposed in Ref.~\cite{Rosenzweig2020}. While the core technology is based on "conventional" RF accelerating structures, several key advancements are proposed, including use of cryo-cooled copper structures to operate at 125 MV/m acceleration gradients in both the photo-injector and the linac. This much increased gradient allows for creation of beams with an order of magnitude higher brightness (20 A peak current and 50 nm emittance from the photoinjector). The increase on linac gradient permits relevant energies near 1 GeV to be reached in 8 meters. To obtain the peak currents needed for FEL lasing the scheme utilizes compact chicanes, a high frequency longitudinal phase space linearizing cavity, and a passive dechirper in combination with inverse-FEL techniques to compress and micro-bunch the electron beam into a series of high-peak-current (4 kA) beamlets. This beam is coupled to short period, mm-scale, undulators of $\sim$5 m length for lasing at soft X-ray energies. Such a system could operate at 100 Hz and fit within a 40 meter footprint including x-ray optics. The foundation of the accelerator technology is maturing,  with acceleration of up to 150 MeV/m shown and the cryogenic gun under construction. Further research is underway to push the peak acceleration gradient past 250 MV/m. Integration of sub-mm-diameter passive dechirpers and other advanced phase-space techniques will need to be explored. However, fundamentally, such a compact FEL, which is being proposed as a development program to both DoE BES and NSF, could be considered for user-based applications in the next 5-10 years. The project, termed UCXFEL, is now seen as a useful stepping stone demonstration project for the C$^3$ cryogenic copper collider which is under intense discussion and will be presented in detail at Snowmass \cite{Nanni2022C3Snowmass}. We also note that there is high interest in UCXFEL/C$^3$ technology in the full scale XFEL community.    
    

An alternative path to a compact x-ray FEL is to accelerate up to ~5 GeV a low-charge, high brightness electron bunch using the wakefield created by a high charge drive bunch in a corrugated waveguide \cite{Zholents2020, SiyPhysRevAccelBeams.25.021302}. Unique features of this path are a high energy efficiency, since up to 80\% of the drive bunch energy is deposited to the wakefield, and a high repetition rate 20 - 50 kHz. The achievable accelerating field is ~100 MV/m.  A 1 GeV  superconducting accelerator operating at 200 - 500 kHz bunch repetition rate can support the array of ten wakefield accelerators and ten XFEL providing a highly flexible operational environment for users of this x-ray facility. A significant effort in the next 5-7 years needs to be applied to get a full-scale demonstration of a corrugated waveguide-based wakefield accelerator which forms the backbone of this FEL concept.

\subsection{Betatron and Compton sources from plasma accelerators}


While sub- and mildly-relativistic electrons from LWFAs ($\sim$1 MeV), which at such energies can already operate at kHz repetition rates, find applications in emerging technologies such as Ultrafast Electron Diffraction imaging \cite{He2013}, this section will focus on conversion of ultra-relativistic electrons to high-energy photons.
The generation of sources of x-rays and gamma-rays from LWFA has the potential to offer complementary alternatives to synchrotrons and other sources based on RF accelerator technologies \cite{Corde_RMP_2013,Albert_PPCF_2016}. The two most mature LWFA-based sources, with applications in imaging for medical and industrial applications, nuclear physics, material and high energy density science already underway, are based on betatron motion and Compton scattering processes \cite{Corde_RMP_2013,Albert_PPCF_2016}. 

When an intense laser ($I>10^{18}$ W/cm$^2$) interacts with an underdense plasma (electron density in the range $10^{17} - 10^{19}$ cm$^{-3}$), the ponderomotive force drives an electron plasma wave that traps electrons and accelerates them to relativistic energies (with a current record at 8 GeV \cite{Gonsalves_PRL_2019}). 
When such electrons are trapped off the main laser axis, or with initial transverse momentum, they also experience transverse restoring forces due to space-charge separation, prompting them to execute transverse oscillations as they are accelerated. 
This betatron motion generates a bright source of x-rays, first observed in 2002 in a beam driven plasma accelerator \cite{Wang_PRL_2002} and in 2004 in a laser-driven one \cite{Rousse_PRL_2004}, with similar properties to those of a synchrotron (a broadband spectrum with photon energies up to a few tens of keV, a collimated beam of a few tens of milliradians, and a micron source size). 
The two additional benefits are a femtosecond-scale pulse duration and the  synchronization between the drive laser beam, the generated particle bunch, and the secondary radiation, which makes Betatron and Compton light sources based on LWFA unique for applications. 

For higher energy photons, Compton scattering, where laser photons are scattered off the LWFA electrons and upshifted to higher energies (up to a factor of 4 times the square of the electron relativistic factor for a head-on collision), is a better alternative. 
The scattered photons can either be from the LWFA-drive beam reflected on a plasma mirror \cite{TaPhuoc_NPho_2012}, or from a second laser beam \cite{Schwoerer_PRL_2006,Powers_Npho_2014}, which was first demonstrated in 2012 and 2006, respectively. A more general discussion on non-plasma and plasma-based Compton sources is presented in sub-section~\ref{icsics}. Note that for a given electron beam energy, Compton scattering will scale to much higher photon energies (the maximum at present is 18 MeV \cite{Sarri_PRL_2014}) than betatron radiation.

Many of the research efforts with betatron and Compton sources are aimed at applications, and thus the development of these sources and their key parameters (energy, photon flux, spatial and temporal resolution) must be done in close coordination with applications in high energy density, biological, planetary, material and astrophysical sciences, and nuclear photonics. 
In terms of source development, the photon flux, overall laser conversion efficiency (currently around $10^{-5}$) and shot-to shot intensity/energy stability need to be improved. 
Some applications, such as x-ray phase contrast imaging of biological objects and laser-driven shocks, as well as time-resolved x-ray absorption spectroscopy, have already been demonstrated. 
These should likely become routine applications for betatron radiation in a near future, where it can be coupled to high power or free electron lasers capable of driving matter to extreme states. 
Other techniques, such as x-ray scattering or diffraction, will require at least 3 orders of magnitude more photons. Compton scattering provides higher photon energies, is easier to tune, and can have a narrower bandwidth (provided the electron energy spread is small and the source operates in the linear regime where the scattering laser intensity is well below $10^{18}$~W/cm$^2$) than betatron radiation. 
Hence, applications are naturally more geared toward gamma-ray radiography, photofission, and possibly nuclear resonance fluorescence. 
Most of these applications have yet to be demonstrated with a LWFA-based Compton source.

\subsubsection{Alternative plasma-based gamma ray sources}

Another approach to generate high energy photons from LWFA is to use the Bremsstrahlung mechanism. Here, the laser beam and the electron beam, as they exit the plasma, collide with a high-Z material (typically Tantalum or Tungsten) to produce a beam of gamma-rays (above MeV photon energies). A high-Z material is used here because the higher-charge nuclei increases the Bremsstrahlung conversion efficiency. As the electrons pass near the nucleus of the material, they are slowed down as their path is deflected. Due to energy conservation, the lost “braking” energy is emitted as a high-energy photon. If the electron comes to a total stop, it transfers all its energy into the photon, therefore the maximum energy of the Gamma-ray beam matches the maximum energy of the electron beam. This configuration has been shown to work operating in the blowout LWFA regime \cite{Glinec_PRL_2005} and in the self-modulated LWFA regime where it benefits from the almost one order of magnitude increased electron beam charge generating a larger number gamma photons \cite{Lemos_PPCF_2018}. It was found experimentally on the Vulcan facility, operating at $5\times 10^{19}$ W/cm$^2$, that the interaction of the laser pulse with a gas jet in front of a tantalum foil enhances the on-axis dose by a factor 5 with respect to interaction on a solid target and produces a gamma-ray lobe directed on-axis with a reduced angular spread \cite{Edwards_APL_2006}.

LWFA Bremsstrahlung sources of MeV Photons have potential applications for non destructive evaluation such as the inspection of cargo containers or complex structures of dense material. These sources present several advantages that are desired for this type of application: high energy photons (MeV) to penetrate dense objects, a low dose for safety, and a small source size (a few microns) for good spatial resolution. The resolution in the first blowout regime experiment was 320 $\mu$m \cite{Glinec_PRL_2005}, which was well improved to 30 $\mu$m with optimization of the electron beam parameters \cite{Benismail_NIMA_2011}. 

\subsubsection{Advanced non-plasma ICS-based gamma ray sources}
\label{icsics}

One potentially promising application of the compact laser-driven 1 GeV class accelerators is Inverse Compton Scattering (ICS) \cite{Sprangle1992, Schoenlein1996}, where the quasi-monochromatic and directional X-rays are generated by colliding energetic electrons with a laser beam. In the ICS process, the excitation of a relativistic electron beam by the counter propagating laser pulse generates relativistically Doppler-shifted forward scattered photons, with the on-axis wavelength of $\lambda_x = \lambda_L/4\gamma^2$, where $\lambda_L$ is the laser wavelength, and $\gamma$ is the Lorentz factor. One can immediately appreciate the tunability of an ICS source due to the quadratic dependence of its wavelength on the electron energy. The laser beam acts as an “optical undulator”, and its micron-scale wavelength makes ICS sources very compact and thus befitting industrial, medical, and field deployable applications. 

However, the ICS cross section is very small (at moderate energies it is the same as the classical Thomson scattering cross section of ~ $6.6\times 10^{-25}$ cm$^2$), thus a well optimized ICS source operating in the linear regime (no significant outcoupling into harmonics) generally generates about $10^6-10^7$ useful photons per interaction, whereas most of the practical applications require on the order of $10^{10}-10^{14}$ photons per second. As a consequence, the ICS source scalability to $>$10 kHz repetition rates and beyond is considered of paramount importance.

The two most interesting practical scenarios to consider are: (1) very compact hard X-ray sources that can bring synchrotron capabilities to users’ facilities (i.e., for medical imaging or nanometrology); and (2) tunable quasi-monochromatic gamma ray sources, which consistent with a footprint of a small laboratory, or even a mobile platform. 
In case of the former, there is no particular benefit of the advanced accelerator ICS driver vs. a conventional linac in terms of the system size (a 1 meter long 30 MeV linac would basically cover the hard X-rays ICS range up to 50 keV), thus unless the advanced system offers some additional application specific benefits (as we shall see for some applications discussed below), a more general area where the  advanced accelerator applications could play a role, are higher energy gamma ray ICS sources.

In that latter case, the benefits of the advanced accelerators become profound.  First of all, in order for the ICS system to operate in the multi-MeV range, a GeV-class electron beam is required, which cannot be produced by the RF linac alone without having to expand the system footprint beyond the practical limits of many key applications. Thus, applying higher gradient advanced acceleration methods in such sources is of the critical value to the potential users. In addition, the laser driven advanced accelerators can in some scenarios recycle its laser beam to drive the ICS interaction, offering very efficient use of the ultrafast laser power, which is always at high premium. In view of that, we narrow our consideration here to the three laser-driven advanced acceleration schemes: Dielectric Laser Accelerator (DLA), Laser Plasma Wakefield Accelerator (LWFAA), and Inverse Free Electron Laser (IFEL).  

The DLA is at the present time the least compatible technology with the gamma-ray ICS source needs, as most of the current work is focused on the lower energy electrons and low intensity beams below the damage threshold of the dielectric optical waveguides. It is important, however, not to rule out the DLA technology entirely, as one day it may achieve the level of maturity consistent with the vision of the mass produced ``accelerator chip", and then find a role to play at some lower energy and lower intensity ICS applications, where the source size miniaturization is required. Nevertheless, we find the prospect of such applications driven DLA-ICS R$\&$D outside of the 5-10 years horizon considered herein.

LWFA on the other hand, has already been demonstrated experimentally as the ICS driver \cite{Powers2013, Kramer2018, Dopp2016} by using a reflective foil at the plasma exit, to reflect the LWFA laser back into the plasma channel and thus excite ICS interaction. Such approach has a number of very promising features.  First, the LWFA laser is directly recycled to produce ICS gamma-rays within the LWFA itself, resulting in a very compact system and economic use of the laser power.  In addition, LWFA ICS can take advantage of the very high gradients achievable by LWFA to produce truly table-top tunable gamma ray sources.  Finally, the electron beam is naturally very small at the IP, due to plasma focusing, resulting in a very high intrinsic source brightness, which is very advantageous feature for some applications (i.e., medical phase contrast imaging at hard X-ray range). Of course, there are also multiple challenges for LWFA ICS technology, such as scalability to very high repetition rates, and control over the ICS output spectral purity. These issues, however, are directly aligned with the broader goals of the LWFA development, thus deploying ICS sources at LWFA facilities could be an important strategy during the next decade, to simultaneously offer gamma rays to the users, and at the same time provide very relevant performance metrics and additional diagnostics to the LWFA developers.

A typical power of the ICS laser to maintain the linear interaction in a tightly focused ICS interaction point as a rule does not exceed TW level, which is significantly below the laser pulsed energy required for a GeV-class laser-plasma source, yet consistent with the laser intensities required to drive Inverse Free Electron Laser (IFEL) \cite{Palmer1972, Courant1985}, a scheme first demonstrated in early 2000s at UCLA and BNL independently \cite{Musumeci1985, Kimura1984}, followed by the groundbreaking UCLA-BNL Rubicon program \cite{Duris2014,Duris2015}. Although IFEL acceleration gradient is modest compared to laser wakefield plasma accelerators (LWFA), it offers a number of important advantages as the GeV-class ICS driver. IFEL is a free space accelerator where the electron beam and laser interact in vacuum in the magnetic field, without energy losses to the media. This allows to reach very high gradients and also very high repetition rates, without risks of material breakdowns or unsustainable heat deposition to the media. Also, in IFEL the part of the laser pulse energy not directly transferred into the energy of the e-beam is well preserved and can be recycled and reused.  Some initial promising experimental results have already been demonstrated \cite{Sudar2020, Gadjev2019}, and there is a technical path to intracavity IFEL-ICS system, which can achieve very high repetition rates within the practical limits of the existing laser technology. Based on these facts, we believe that IFEL-ICS source technology offers a promising and practical path towards compact and tunable high flux multi-MeV gamma ray sources for nuclear science and applications, and this technology should be (and will be) explored experimentally over the next 5-10 years period.

\section{Medical applications driven by particle beams}

\subsection{Laser-based ion acceleration}

The development of ion beams for tumor therapy has contributed significantly to advancing the field of radiotherapy towards curative treatments \cite{Blakely2020}. External high-energy ion therapy benefits from the inverse depth dose profile of ion stopping in tissue, in the so-called Bragg peak, where the ion dose is primarily deposited in a narrow range at depth, thus sparing surrounding normal tissue in front of and behind the tumor volume. Access to ion therapy facilities is limited due to their size and operation cost, which effectively limit world-wide patient access to this type of treatment. Laser-driven (LD) ion sources are receiving increasing attention due to their potential of generating proton and heavier ion beams for radiotherapy on a relatively small footprint compared to conventional (radio frequency-driven) particle therapy facilities \cite{Bulanov2004,Ledingham2014,Malka2004}. 
However, stringent requirements concerning combined key beam parameters like proton energy (up to 250 MeV), numbers of protons per bunch (10$^9$), stability and control of energy and proton numbers from shot to shot ($<$ few percent) and repetition rate ($>$ 10 Hz) are yet to be demonstrated in experiments \cite{Bolton2016,Linz2016,Enghardt2018}. 
It should also be noted that more compact conventional proton machines have been actively pursued with impressing results\cite{Contreras2017}, although no similar effort is underway for heavier ion machines. \newline

At the current state, reaching clinically relevant particle energies represents a primary challenge for the field of laser-ion acceleration that is expected to see significant advances from the ongoing efforts to develop high-repetition rate, several PW-class lasers.
So far, highest proton energies achieved in experiments are approaching 100 MeV \cite{Higginson2018,Kim2016PoP}, which is well below energies necessary for clinically relevant penetration depths of $>$30 cm in humans \cite{Enghardt2018}. However, gantry designs and treatment planning studies have been developed for laser-driven ion beam therapy \cite{Masood2017,Hofmann2015}. 
Advances in laser technology are expected to deliver higher LD proton and ion energies because experiments and simulations have shown a consistent increase of maximum particle energies with laser pulse energy, power, or intensity \cite{Macchi2013, Daido2012}.
At the same time, theory and simulations predict higher proton and ion energies when harnessing advanced acceleration regimes like radiation pressure acceleration \cite{Esirkepov2004}, magnetic vortex acceleration \cite{Park2019} or shock acceleration \cite{Fiuza2013}, to name a few. Review papers like references \cite{Macchi2013,Daido2012} provide summaries of previous works in this regard.\newline

\begin{figure}[ht]
    \centering
    \includegraphics[width=0.95\columnwidth]{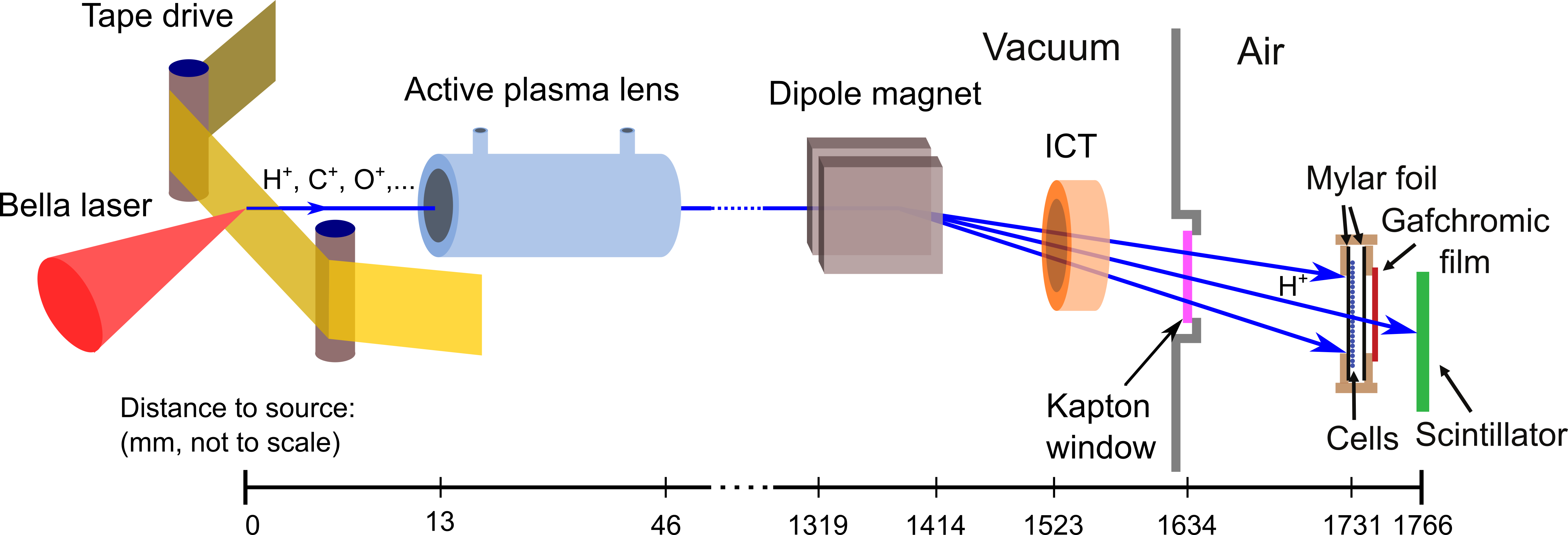}
    \caption{Figure adapted from Ref. \cite{Bin2022}. Schematic depiction of the laser-driven proton beamline at the BELLA PW laser.}
    \label{fig:bella_beamline}
\end{figure}

While clearly more work needs to be done before LD proton sources can be credibly advanced into medical machines, they may soon become adequate complements to conventional accelerators for basic radiobiological research into the effects of particle radiation on cells, tissues and small animals that are ultimately relevant for particle therapy \cite{Friedl2018}.
Access to conventional experimental and medical machines has been rather limited for this type of research \cite{Durante2014} while the steady increase in available compact LD proton sources has already started to open up new experimental options for systematic studies.
As such, an appreciable number of \textit{in vitro} studies has been conducted to investigate the radiobiological effectiveness of LD protons \cite{Yogo2009,Kraft2010,Bin2012,Doria2012,Raschke2016,Manti2017,Bayart2019,Hanton2019}. Proton beamlines have been implemented at a few laser facilities to transport LD protons and shape a three-dimensional dose profile for \textit{in vivo} studies with small animals \cite{Pommarel2017,Brack2020,Cirrone2020}.
Moreover, due to their uniquely short bunch length of $<$ ps at the source, LD proton pulses are valuable tools to investigate the FLASH radiotherapy effect, observed in irradiation studies with ultra-high dose rates \cite{Favaudon2014}.
Recently re-visited after decades of anecdotal reports, the FLASH effect describes the beneficial differential effects on tumors versus normal tissues using the delivery of high radiation doses at extremely high dose rates ($>$40 Gy/s with doses $>$10 Gy delivered in $<$100 ms)\cite{Vozenin2019,Friedl2021}. LD proton and heavier ion pulses can deliver several orders of magnitude higher instantaneous dose rates (IDR) than typical conventional (radiofrequency) accelerators, potentially further increasing the differential sparing effect on normal tissue and consequently broadening the therapeutic window for radiotherapy.\newline

With the biological mechanisms underlying the FLASH effect remaining largely unknown, research into the radiobiological effects of ultra-high dose rate proton irradiation has been hampered by limited access to conventional accelerators that are able to provide the required dose rates \cite{Grilj2020,Hughes2020}. 
In a preliminary study at the 40 Joule BELLA petawatt laser proton beamline, it was demonstrated for the first time that LD protons delivered at ultra-high IDR can indeed induce the differential sparing of normal versus tumor cells \textit{in vitro} for total doses $\geq$ 7 Gy \cite{Bin2022}. In that study, normal and tumor prostate cells in 1 cm diameter cups were irradiation with LD protons of 2-8 MeV at an IDR of 10$^7$ Gy/s. After acceleration from a tape drive target the proton bunch was transported with a compact active plasma lens beamline \cite{Tilborg2015} to the cell sample site located outside the vacuum chamber (Fig. \ref{fig:bella_beamline}). With 1 Gy applied per laser shot, total dose values up to $>$ 30 Gy were accumulated by operating the LD proton beamline at 0.2 Hz. It should be pointed out that the generation of LD proton beams in the energy range sufficient for this type of study does not require a PW laser system but was demonstrated in numerous experiments at 100 TW-class laser systems \cite{Macchi2013,Daido2012}.\newline

To summarize, while current LD ion source parameters are well below the requirements for their use in external ion beam therapy, their comparatively low-cost and compact nature has earned LD ion sources increasing attention for their potential as future compact therapy machines. Already at less stringent source requirements, they could complement conventional accelerators to increase and democratize access to ion sources for pre-clinical radiobiological research. In particular, first studies have demonstrated their capability of delivering ultra-high dose rate proton beams enabling research into the FLASH radiotherapy effect, which could change the way radiotherapy is delivered in the future.    

\subsection{Applications of Dielectric Laser Accelerators}

With tabletop DLA sources now coming into operation in university labs, near-term applications that utilize presently available low-current beams with moderate particle energies in the 100 keV to few MeV range are being actively pursued. Due to the intrinsic optical-scale bunch structure, with sub-femtosecond bunch duration, compact DLA electron sources for ultrafast science and electron diffraction studies are among the most promising applications. Compact accelerators with target energies in the few MeV range for medical dosimetry also provide a compelling near-term use for DLA technology. The sub-femtosecond high brightness electron sources based on laser-driven on-chip accelerators could potentially be fit on the end of an optical fiber, placed on a scanning platform at the surface of a sample, or even inserted into living tissue, to supply unprecedented time, space, and species resolution.

\begin{figure}
\begin{center}
 \includegraphics[height=.18\textheight]{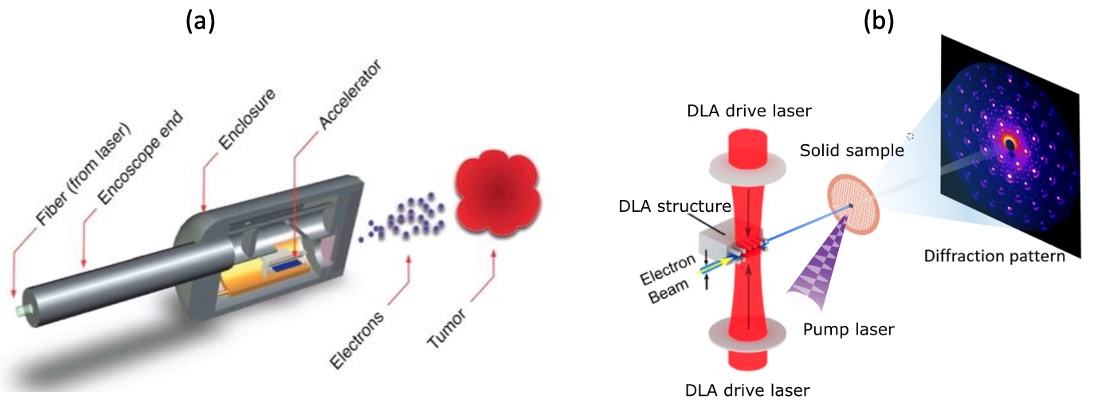}
\caption{Promising near-term applications for laser-driven accelerators include (a) endoscopic medical radiation therapy delivery devices, (b) tabletop sources for MeV ultrafast electron diffraction studies.}
\label{applications}
\end{center}
 \end{figure}

In the realm of medical accelerators, an ultracompact, self-contained multi-MeV electron source based on integrated photonic particle accelerators could enable minimally invasive cancer treatments and adjustable dose deposition in real-time, with improved dose control. For example, one could envision an encapsulated micro-accelerator built onto the end of a fiber-optic catheter placed within a tumor site using standard endoscopic methods, allowing a doctor to deliver the same or higher radiation dose to what is provided by existing external beam technologies, with less damage to surrounding tissue \cite{england:aac14, DOE:CASM:2020}. As electrons have an energy loss rate of about 2 MeV/cm in water, their irradiation volumes can be tightly controlled. Encapsulated devices would ideally have variable electron energies in the 1-10 MeV range, a footprint that is millimeter-scale, and accommodate a wide range of emission angles for various treatment modalities. The manufacturing and operating costs based on low-cost or disposable DLAs, powered by an external fiber laser, could be much lower than those for conventional radiation therapy machines, and the robustness of such systems compared to conventional accelerators could be even more favorable.

Pulse trains of attosecond electron bunches are intrinsic to the DLA approach and could provide excellent probes of transient molecular electronic structure. Recently, this fine time-scale structure has been experimentally measured (with bunch durations from 270 to 700 attoseconds) and injected into a subsequent acceleration stage to perform fully on-chip bunching and net acceleration demonstrations \cite{black:Atto:2019,schoenenberger:Atto:2019}. At present, attosecond electron probes are limited to strong-field experiments where laser field-ionized electrons re-scatter from their parent ion. The far greater flexibility of on-chip attosecond electron sources could be important complements to optical attosecond probes based on high harmonics or attosecond X-ray free electron lasers (FELs). The capabilities of chip-scale electron probes could be transformational for studies of chemical impurities and dopants in materials. Many nanoscale technologies operate by inter- and intra-molecular processes with intrinsic timescales measured in femtoseconds; proton migration in materials occurs over tens of femtoseconds, and electron migration over single femtoseconds. Examples include quantum memory storage and switching using atomic impurities in pure crystalline hosts and lab-on-a-chip technologies for chip-scale genome decoding. DLA technology could also enable a unique class of compact tabletop electron sources for ultrafast electron microscopy (UEM). Since both the accelerator and associated drive laser and peripherals are all compact enough to fit on a tabletop, this would enable the construction of compact and flexible attosecond electron and photon sources that can resolve atomic vibrational states in crystalline solids in both spatial and temporal domains simultaneously. This could lead to improved understanding of the dynamics of chemical reactions at the level of individual molecules, the dynamics of condensed matter systems, and photonic control of collective behaviors and emergent phenomena in quantum systems.

The difference in bunch charge and duration for DLA electron beams also points to the potential for future DLA-based light sources for generation of attosecond-scale pulses of extreme ultraviolet (EUV) or X-ray radiation. The DLA approach has the potential to produce extremely bright electron beams that are suitable for driving superradiant EUV light in a similarly optical-scale laser-induced undulator field \cite{england:FLS2018}. Furthermore, because of the few-femtosecond optical cycle of near infrared mode-locked lasers, laser-driven undulators could potentially generate attosecond X-ray pulses to probe matter on even shorter time scales than is possible today. Laser-driven dielectric undulators have been proposed and could be fabricated using similar photolithographic methods used to make on-chip accelerators \cite{plettner:2008,plettner:2009}. Combining the high gradient and high brightness of advanced accelerators with novel undulator designs could enable laboratory-scale demonstrations of key concepts needed for future EUV and X-ray lasers that hold the potential to transform the landscape of ultrasmall and ultrafast sciences. To realize these laboratory-scale, lower-cost, higher performance radiation sources, critical components of laser-driven radiation sources need to be developed and demonstrated. Coherent attosecond radiation could potentially be produced using the same operating principles that produce particle acceleration via the “accelerator on a chip" mechanism. These structures operate optimally with optical-scale pulse formats, making megahertz repetition rate attosecond-scale pulses a natural combination. The optical-scale undulator regime has not been extensively studied before and questions arise as to how well the beam will behave in such structures and how well it will ultimately perform. The theoretical and numerical tools to model these processes need to be developed in order to guide experimental studies of attosecond electron and photon generation.

Various technical challenges to developing near-term applications for tabletop laser accelerators (including high-energy gain, multi-stage beam transport, and higher average beam power) must also be addressed in order to reach the stringent beam quality and machine requirements for a HEP collider. The near-term applications outlined above therefore represent an intermediate advancement of the technology towards ultimately desired HEP performance levels. Further development of these applications also holds the potential to leverage both scientific and industrial interests that could facilitate more rapid development and support.

\subsection{Laser-plasma-accelerator based VHEE (very high electron energy) radiation therapy}

Laser-plasma accelerators are currently deserving a great attention in view of possible novel radiotherapy (RT) protocols, based on the direct usage of electron beams.
Currently, in the overwhelming majority of the cases, RT is delivered using Brems\-strahlung photons from $\sim 10 \,\mathrm{MeV}$ electrons. 
The direct usage of electron beams has been hampered so far by the limited electron energy from reasonably compact accelerators. 
As a matter of fact, electron beams of a few tens of MeV energy suffer from major weaknesses as their dose delivery properties in human tissues are considered; indeed, they features, for instance, a limited depth range, important transverse scattering, large penumbra, relatively high range straggling. 
These issues ultimately makes the dose deposition pattern from electrons with energy in this range of a rather poorer quality compared to that achievable using photons.
This well consolidated scenario is susceptible to be dramatically changed if the usage of so-called Very High Energy Electrons (VHEEs), with energy in the range $\sim 100 - 250\,\mathrm{MeV}$, is considered.
With this respect, LWFA offers the most promising method for compact and affordable VHEE accelerators to become available in the medical environment.
This fact has motivated, over the past few years, a renewed interest in the study of the dosimetric properties of VHEE beams and of their potential for novel RT protocols.
Beside overcoming most of the shortcomings cited above, RT based on VHEEs would offer some more appealing features, such as the possibility of fast steering of the beam to possibly follow the respiratory motion or its focusing to locally enhance the dose on millimeter-sized regions.
Very recently, experiments aimed at assessing the feasibility of advanced irradiations of interest for novel RT protocols have been reported \cite{Svendsen2021,Labate2020,Kokurewicz2019}.
With this respect, one of the issues still to be addressed for a translation into the clinical practice pertains to the long term stability and reproducibility, with the figures typical of medical applications, of LWFA based devices.\par
One of the most relevant cases for the usage of laser-driven electron accelerators in RT can actually come from the recent discovery of the so-called FLASH effect \cite{Favaudon_SciTranslMed2014}, basically consisting in a surprising and remarkable sparing (in terms of unwanted damage of different kinds) of healthy tissues, at a given dose, when dose is delivered at ultrahigh dose rates \cite{Wilson_FrontOnc2020}.
While the radiobiological mechanisms underpinning this effect is still unclear and the subject of a very active field of study, there is a common awareness that this effect may possibly revolutionize the RT practice.
With this respect, while conventional LINAC based machines are being developed \cite{Maxim_RadiotOncol2019}, LWFA may offer one of the most attractive routes to compact FLASH machines, with a potential widespread use \cite{Durante_BrJRadiol2018}.

\section{Fundamental applications}

\subsection{Particle physics beam dump experiments}

\begin{figure}[ht]
    \centering
    \includegraphics[width=0.95\columnwidth]{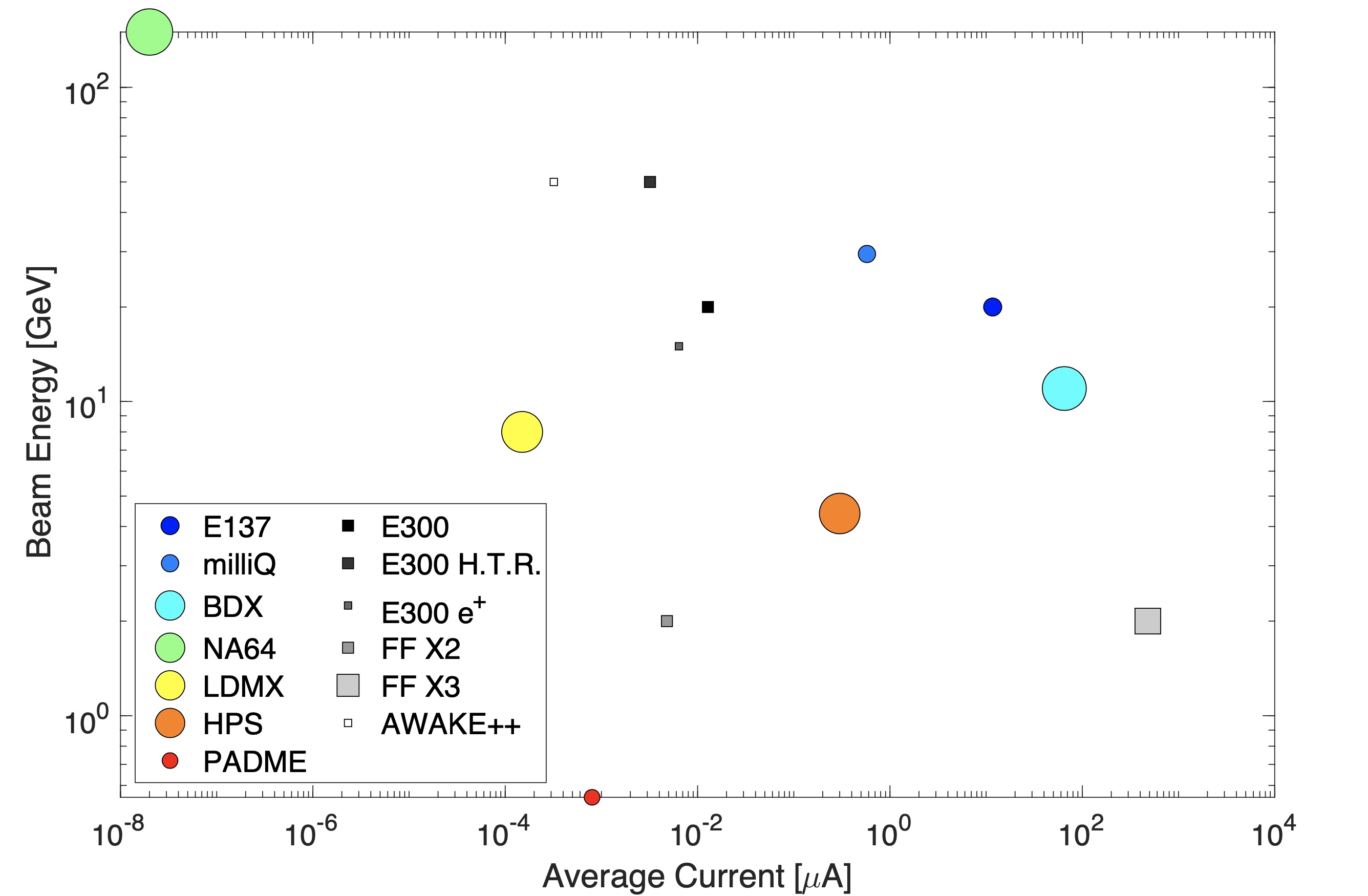}
    \caption{Colored circles: Beam energy and average current for past and planned beamdump experiments. The size of the circle is proportional to the log of the beam rate. Gray squares: Beam energy and average current of planned plasma acceleration experiments. The size of the square is proportional to the log of the beam rate. Plot reproduced from Ref.~\cite{GessnerLOI}}
    \label{fig:current_v_energy}
\end{figure}

Beamdump experiments employ a variety of techniques to detect rare interactions. For the purposes of this paper, the most significant difference between the experiments is whether or not the search relies on kinematic measurements of beam particles. For kinematic searches, the detector attempts to reconstruct individual particle trajectories and is therefore limited to low average currents on target, but at very high repetition rate~\cite{Banerjee2020,LDMX,PADME}. Other experiments look directly for particles that are generated in and pass through the beam dump before reaching the detector~\cite{Bjorken1988,Prinz1998,Battaglieri2016}. These experiments expect high average current and are compatible with high-charge bunched beams. Some detector types work best with bunched beams because this is useful for rejecting out-of-time backgrounds~\cite{SnowdenIfft2019}. Finally, positron beams may be used to enhance production rates of dark matter particles and are compatible with a bunched-beam format~\cite{Marsicano2018}.

Figure~\ref{fig:current_v_energy} plots the beam energy and average current of past and planned beam dump experiments log-log scale. The log of the repetition rate of the experiments is represented by the size of the marker. Parameters of planned plasma wakefield acceleration experiments are shown on the same plot. Results are expected from the E300 experiment at FACET-II and the X2 experiment at FlashForward within the next two years. Results from the E300 high-transformer ratio experiment, E300 positron experiment, and X3 experiment are expected within five years. The AWAKE++ experiment will follow the currently planned AWAKE Run-II experiment which aims to demonstrate 10 GeV acceleration in the next 5 years. Plasma wakefield accelerators may therefore provide beams relevant for these experiments in the coming decade.

\subsection{Studies of astrophysical plasmas}

Relativistic plasma processes shape the dynamics, energy partition, and multi-messenger signatures of some of the most fascinating astrophysical environments --- from the plasmas orbiting black holes and neutron stars to gamma-ray bursts (GRBs) and relativistic jets from active galactic nuclei (AGN). Of primary importance to many of these physical systems are the so-called streaming instabilities, which arise in plasmas hosting interpenetrating flows --- a common configuration in high-energy astrophysical and laboratory plasma environments. Relativistic streaming instabilities develop when the relative velocity between the charge-particle beams and the ambient medium approaches the speed of light and constitute a long-standing problem in fundamental plasma physics. 
These instabilities control the energy transfer between relativistic flows, fields, and particles in these environments. However, because the properties of these relativistic plasmas can depart significantly from those of traditional non-relativistic plasmas studied in the laboratory and in space, our understanding of the basic plasma processes in this frontier regime is still very limited.

The physics of relativistic plasmas and their underlying instabilities have been mainly
explored through analytical theory and kinetic simulations. Previous studies have established plasma microinstabilities, such as the oblique/two-stream (electrostatic) \cite{bret10} and the Weibel/current-filamentation (electromagnetic) \cite{weibel59,Fried59} as key players in the generation of electromagnetic turbulence \cite{Kazimura1998,silva03}, particle acceleration \cite{Spitkovsky-2008,sironi15,Lemoine19}, and radiation emission \cite{Medvedev_Loeb_1999,Gruzinov_Waxman_1999} in relativistic collisionless shocks, such as in GRBs. They can also be important in the transport of highly relativistic pair beams in AGN jets \cite{broderick12,sironi14} and may account for the origin of cosmic magnetic fields on various scales \cite{schlickeiser05}. Despite the promising advances in kinetic simulations, understanding the competition between the different instabilities, the long-term nonlinear evolution, and the true saturation behavior, both in terms of field strength and coherence length, remains a fundamental challenge.

Advances in accelerator technology are creating unique opportunities to probe relevant relativistic plasma instabilities in controlled laboratory experiments for the first time. In particular, tightly focused, relativistic electron beams \cite{Yakimenko19} can trigger the fast growth of relativistic plasma microinstabilities and enable detailed studies of their properties and nonlinear evolution that can be used to benchmark theoretical and numerical models. As the beam propagates in the plasma, these instabilities will exponentially amplify self-generated electromagnetic fields. The violent acceleration experienced by the beam electrons in these strong fields leads to significant synchrotron emission. The resulting high-energy photon pulse has a duration comparable to that of the electron beam and is highly collimated, providing an important signature of the underlying processes. Indeed, the same plasma instabilities 
are thought to mediate the amplification of magnetic fields, the slow down of highly-relativistic plasma flows, and the high-energy radiation emission in these extreme astrophysical settings \cite{Medvedev_Loeb_1999,Gruzinov_Waxman_1999}. 

Moreover, when the beam density is comparable to the background plasma density, the conversion efficiency between the electron beam and the high-energy photons can largely exceed 10 $\%$, potentially enabling the development of high-brilliance gamma-ray light sources based on the plasma instabilities. For example, it is expected that dense (nC, few $\mu$m size) multi-GeV electron beams can produce collimated gamma-ray pulses with peak brilliance above $10^{25}$ photons s$^{-1}$ mrad$^{-2}$ mm$^{-2}$ per $0.1\%$ bandwidth in the MeV to GeV range \cite{benedetti18}. This research could drive the development of stable high-energy, high-density beams for a future light source facility that would also benefit designs for short bunch particle colliders for high-energy physics studies.

The experimental study of relativistic plasma instabilities using high-energy particle beams will strengthen the connection between advanced accelerator technology, fundamental plasma science, and phenomenological models for high-energy astrophysical plasmas with far-reaching implications for our understanding of some of the most extraordinary phenomena in the Universe. It will also establish unique opportunities to control and explore these basic processes in the laboratory and
to validate numerical models commonly used in laboratory and astrophysical plasma research for unprecedented conditions.

\section{Conclusion}

In this white paper we have summarized and discussed the ideas presented by the advanced accelerator community for near term applications of advanced accelerator technology. We have emphasized the scientific and societal impact of these applications as well as their role in the context of improving advanced accelerator performance and reliability on the path towards an advanced accelerator linear collider. A strong and vibrant community pursuing the near term applications discussed in this manuscript is poised to benefit the high energy physics community going forward. It is therefore important that such applications are nurtured in an ecosystem that affords them the resources needed to succeed, both in terms of dedicated economic support and access to facilities appropriate for developing near term applications and their associated technologies. 


\bibliographystyle{unsrt}
\bibliography{main.bib}

\end{document}